\documentclass[letterpaper,english,aip,showpacs,showkeys]{revtex4-1}

\usepackage[T1]{fontenc}
\usepackage[utf8]{inputenc}
\usepackage{amsmath}
\usepackage{amssymb}
\usepackage{babel}



\begin{document}

\title{Eigensolutions of the two Dimensional Kemmer Oscillator in Noncommutative Space with Minimal Length Effects}

\author{Abdelmalek Boumali}
\email{boumali.abdelmalek@gmail.com}
\affiliation{Laboratoire de Physique Appliquée et Théorique, \\
Université Larbi-Tébessi- Tébessa, Algeria}

\begin{abstract}
{\normalsize Abstract: This paper delves into the intricacies surrounding
a two-dimensional Kemmer oscillator within the domain of relativistic
quantum mechanics, enriched by the incorporation of minimal length
and the integration of non-commutative phase space effects. Our investigation
yields the derivation of eigen solutions, shedding light on the intricate
interplay between minimal length and non-commutative parameters. These
solutions are derived within the configuration space $\{p\}$. Through
a thorough analysis, we establish a compelling connection with the
Schrödinger equation, which manifests itself in the form of a Pöschl-Teller
potential. This exploration not only unveils the nuanced behavior
of the Kemmer oscillator within an innovative theoretical framework
but also deepens our understanding of fundamental quantum phenomena
in relativistic systems.}
\end{abstract}

\maketitle

\section{Introduction}

In the realm of relativistic quantum mechanics, the pursuit of precise
solutions to wave functions holds profound significance, as it reveals
the intricate underlying physics. The origins of the relativistic
wave function for a massive spin-1 particle can be traced back to
Kemmer's seminal work in 1939 \cite{1,2}. Known as the Kemmer equation,
it presents a Dirac-type formulation characterized by matrices governed
by unique commutation rules. Notably, our focus lies on the scenario
of a massive spin-1 particle, which comprises a dual-particle system
involving spin-1/2 constituents, rather than solely a single spin-1
particle. This distinctive characteristic endows the Kemmer equation
with the attributes of a dual-body Dirac-like equation, stirring recent
heightened interest, particularly regarding its potential to elucidate
the nature of quark-anti-quark bound states \cite{1,2,3,4,5,6}.

The Dirac relativistic oscillator holds significant theoretical and
practical importance. Initially investigated by Ito et al. \cite{7},
the Dirac equation was formulated with the momentum $\vec{p}$ replaced
by $\vec{p}-im\beta\omega\vec{r}$, where $\vec{r}$ denotes the position
vector, $m$ represents the particle's mass, and $\omega$ indicates
the oscillator frequency. Renewed interest in this phenomenon was
sparked by Moshinsky and Szczepaniak \cite{8}, who termed it the
Dirac oscillator (DO). In the non-relativistic limit, it behaves as
a harmonic oscillator with a robust spin-orbit coupling term.

Physically, the interaction of the DO can be understood as the anomalous
magnetic moment interacting with a linear electric field\cite{9,10}.
Benitez et al. \cite{11} delineated the associated electromagnetic
potential. The Dirac oscillator has garnered substantial attention
as one of the exact solvable examples of Dirac's equation, with numerous
physical applications (see Ref. \cite{6} and associated references).
Franco-Villafane et al. \cite{12} recently outlined the initial
experimental realization of the one-dimensional DO using microwaves.
The experiment exploits the connection between the DO and a corresponding
tight-binding system. Results obtained from the experiment closely
align with theoretical predictions concerning the spectrum of the
one-dimensional DO with and without the mass term.

The Dirac oscillator remains a subject of active research interest
among physicists aiming to unravel the intricate connections between
relativity theory and quantum mechanics. An intriguing example where
the Dirac oscillator finds relevance is in the study of graphene.
Graphene, a single layer of carbon atoms arranged in a hexagonal lattice,
exhibits unique electronic properties due to its two-dimensional structure
and the presence of Dirac cones in its band structure. In graphene,
electrons behave as massless relativistic particles described by the
Dirac equation near the Dirac points. When graphene is subjected to
an external periodic potential, the resulting system can be effectively
modeled as a Dirac oscillator{[}{]}.

The exploration of non-commutative (NC) spaces and their implications
in physics constitutes a dynamic and burgeoning field of research.
Many theorists contend that non-commutativity should be regarded as
an inherent feature of spacetime, particularly at the Planck scale.
Simultaneously, the study of quantum systems within NC spaces has
garnered significant attention, driven by the hypothesis that non-commutativity
may be a consequence of quantum gravity effects. In these investigations,
a special focus has been placed on non-commutative quantum mechanics
(NCQM) models . This approach carries intrinsic value as NCQM serves
as a fertile theoretical playground where we can glean insights into
the repercussions of non-commutativity in field theory, employing
conventional quantum mechanics calculation techniques. Numerous NC
field theory models and extensions of quantum mechanics have been
examined, with phase space non-commutativity emerging as a particularly
compelling domain. Phase space non-commutativity has found relevance
in the realms of quantum cosmology, black hole physics, and the conundrum
of singularities. Furthermore, this specific formulation proves indispensable
for incorporating Bose–Einstein statistics within the context of NCQM.
(see \cite{13,14,15,16,17,18,19,20,21}).

The quest for reconciling the general theory of relativity with quantum
mechanics stands as one of the most pivotal challenges in theoretical
physics. This endeavor anticipates the emergence of a fundamental
limit to length measurement, often conjectured to align with the Planck
length scale. Across various quantum gravity frameworks, there exists
a consensus that near this Planck scale, the conventional Heisenberg
uncertainty principle necessitates a reevaluation. The concept of
a minimal measurable length arises notably within string theory, wherein
it stems from the inherent limitation that a string cannot probe distances
smaller than its characteristic scale, $\hbar\sqrt{\beta}$, with
$\beta$ representing a minute positive parameter known as the deformation
parameter. This minimal length adjustment introduces an additional
layer of uncertainty to position measurements, thereby prompting a
revision of the canonical commutation relation between position and
momentum operators.
\begin{equation}
\left[\hat{x},\hat{p}\right]=i\hbar\left(1+\beta p^{2}\right)\label{eq1}
\end{equation}
This commutation relation leads to the standard Heisenberg uncertainty
relation 
\begin{equation}
\Delta\hat{x}\Delta\hat{p}\geq i\hbar\left(1+\beta\left(\Delta p\right)^{2}\right)\label{eq2}
\end{equation}
This clearly suggests the presence of a non-zero minimal length, denoted
as $\triangle x_{\text{min}}=\hbar\sqrt{\beta}\sim l_{p}$, where
$l_{p}$ represents the Planck length. This adjustment to the uncertainty
relation is commonly referred to as the generalized uncertainty principle
(GUP) or the minimal length uncertainty principle \cite{22,23,24,25}.
It's noteworthy that Saavedra and Utreras \cite{26} were the pioneers
in proposing a modification of the canonical commutation relations
of quantum mechanics, anticipated to hold significance at elevated
energy levels. Consequently, a novel (high energy) uncertainty principle
emerged as a result.

In this context, several observations can be drawn regarding Eqs.
(\ref{eq1}) and (\ref{eq2}): (i) Notably, as highlighted in the
works of \cite{23,24,25}, it has been recognized that several observable
effects stemming from the minimal length uncertainty relation exhibit
non-perturbative behavior with respect to the deformation parameter
$\beta$. This implies that these effects encompass contributions
from all orders in $\beta$, despite the fact that $\beta$ appears
only linearly in Eqs. (\ref{eq1}) and (\ref{eq2}). (ii) In the investigation
conducted in Ref. \cite{27,28}, the authors delve into the impact
of the minimal length on the thermal characteristics of a Dirac oscillator,
wherein the positions and momenta adhere to Eq. (\ref{eq1}). They
explore potential constraints on $\beta$, ultimately determining,
through the utilization of properties of the Epstein zeta function,
a minimal length range within $10^{-16}<\Delta x<10^{-14}\text{m}$,
while imposing the physically viable condition $\beta>\beta_{0}=\frac{1}{m^{2}c^{2}}$.
(iii) Kempf \cite{23,24,25} elucidated that Eq. (\ref{eq1}) naturally
emerges when modifying the canonical commutation relation between
position and momentum operators in accordance with Eq. (\ref{eq1}).
Consequently, the choice of position $\hat{x}$ and momentum $\hat{p}$
operators may be made with reference to the undeformed observables
$x$ and $p$, where $[\ensuremath{x,p}=\ensuremath{i\hbar}]$
\begin{equation}
\hat{x}=\left(1+\beta p^{2}\right)x,\quad\hat{p}=p\label{eq3}
\end{equation}
One should note that the construction of models in these spaces would
not be an easy task as the operators $\hat{x}$ and $\hat{p}$ in
the deformed commutation relation (\ref{eq1}) are in general not
Hermitian $\hat{x}^{\dagger}=\hat{x}+2i\beta\hbar\hat{p}$ and $\hat{p}^{\dagger}=\hat{p}$,
albeit the simplified version (\ref{eq3}) still allows one operator
to remain Hermitian \cite{29,30,31,32,33}. According to the works
of Bagchia et al.\cite{34} the positivity of $\beta$ becomes important,
as it ensures the absence of singularities in the metric.

According to the Kempf's prescription, the position and momentum operators
satisfying equation (\ref{eq1}) can be also represented by 
\begin{equation}
\hat{x}=i\hbar\left(1+\beta p^{2}\right)x+\gamma p;\qquad\hat{p}=p,\label{eq4}
\end{equation}
where the operators $x$ and $p$ satisfy the canonical commutation
relation$[x,p]=i\hbar$. In this case, the internal product in Fourier
space becomes 
\begin{equation}
\left\langle \phi\left(p\right)^{*}\psi\left(p\right)\right\rangle =\int\frac{\phi\left(p\right)^{*}\psi\left(p\right)}{\left(1+\beta p^{2}\right)^{1-\frac{\gamma}{\beta}}}dp.\label{eq5}
\end{equation}
The parameter $\gamma$ appears in both Eqs. (\ref{eq4}) and (\ref{eq5})
is an arbitrary constant which does not affect the observable quantities;
its choice determines only the weight function in the definition of
the scalar product \cite{35}. In this work, we have opted with the
Kempf method, and so we chose $\gamma=0$. Nowadays, the reconsideration
of the relativistic quantum mechanics in the presence of a minimal
measurable length have been studied extensively. In this context,
many papers were published where a different quantum system in space
with Heisenberg algebra was studied (see the following references
\cite{30,31,32,33,35,36,37,38,39,40,41,42,43,44}.

The aim of this study is to explore the construction of a two-dimensional
Kemmer oscillator within the context of relativistic quantum mechanics,
incorporating minimal length considerations in NC space. Initially,
we translate the problem into a commutative space through suitable
transformations before addressing it under the influence of a minimal
length.

The paper's organization is outlined as follows: Section II addresses
the problem within the NC space, exploring its examination within
the framework of relativistic quantum mechanics and considering the
effects of minimal length. Finally, Section III presents our conclusions.

\section{The solutions of the Kemmer oscillator in NC space in the presence
of a minimal length}

\subsection{Two dimensional Kemmer oscillator in commutative space}

The relativistic Kemmer equation for spin-1 particles, akin to the
Dirac equation, is expressed as follows {[}1, 2{]}:
\begin{equation}
\left(\beta^{\mu}p_{\mu}-Mc\right)\psi_{K}=0\label{eq:6}
\end{equation}
Here, $M$ represents the total mass of two identical spin-1/2 particles.
The $16\times16$ Kemmer matrices $\ensuremath{\beta^{\mu}}(\ensuremath{\mu=0,1,2,3})$
adhere to the relation:
\begin{equation}
\beta^{\mu}\beta^{\nu}\beta^{\lambda}+\beta^{\lambda}\beta^{\nu}\beta^{\mu}=g^{\mu\nu}\beta^{\lambda}+g^{\lambda\nu}\beta^{\mu}\label{eq:7}
\end{equation}
where $\beta^{\mu}=\gamma^{\mu}\otimes I+I\otimes\gamma^{\mu}$. Here,
$I$ represents a $4\times4$ identity matrix, $\gamma^{\mu}$ are
Dirac matrices, and $\otimes$ denotes a direct product. The stationary
state $\psi_{k}$ of equation (\ref{eq:6}) is delineated as a four-component
wave function:
\begin{equation}
\psi_{K}=\psi_{D}\otimes\psi_{D}=\begin{pmatrix}\psi_{1}\\
\psi_{2}
\end{pmatrix}\otimes\begin{pmatrix}\psi_{1}\\
\psi_{2}
\end{pmatrix}=\left(\psi_{1},\psi_{2},\psi_{3},\psi_{4}\right)^{T}\label{eq:8}
\end{equation}
where $\psi_{D}$ denotes the wave function of the Dirac equation.

In the presence of the Dirac oscillator potential, the momentum operator
$\boldsymbol{p}$ in the free Kemmer equation can be replaced by $\boldsymbol{p}-iMB\omega\boldsymbol{x}$,
where the additional term is linear in $|x|$. Consequently, the Kemmer
equation with a Dirac oscillator interaction becomes:
\begin{equation}
\left[\left(\gamma^{0}\otimes I+I\otimes\gamma^{0}\right)E-c\left(\gamma^{0}\otimes\boldsymbol{\alpha}+\boldsymbol{\alpha}\otimes\gamma^{0}\right)\left(\boldsymbol{p}-iMB\omega\boldsymbol{x}\right)-Mc^{2}\left(\gamma^{0}\otimes\gamma^{0}\right)\right]\psi_{K}=0\label{eq:9}
\end{equation}
In $(1+2)$ dimensions, the standard Dirac $\gamma$ matrices are
substituted by Pauli $\sigma$ matrices, where $\omega$ denotes the
oscillator frequency, and the operator $B$ is defined as $\ensuremath{B=\gamma^{0}\otimes\gamma^{0}},\text{with}\ensuremath{B^{2}=I}$.
Here, the form of the operator $B$ has been proposed based on the
following rationale: beginning with the Dirac oscillator (DO), we
assert that $\ensuremath{\boldsymbol{p}\rightarrow\boldsymbol{p}-im\omega\gamma^{0}\boldsymbol{r}},\text{where}\ensuremath{(\gamma^{0})^{2}=I}$.
Since our system comprises two subsystems of spin-1/2 particles, we
choose an operator $B$ formed as the tensor product of two matrices
$\gamma^{0}$ to ensure the condition $B^{2}=I$. This approach is
elucidated in References 6 and 8.

Additionally, $\boldsymbol{\alpha}=\gamma^{0}\boldsymbol{\gamma}$,
where 
\begin{equation}
\ensuremath{\alpha_{x}=\sigma_{x}=\begin{pmatrix}0 & 1\\
1 & 0
\end{pmatrix}},\quad\ensuremath{\alpha_{y}=\sigma_{y}=\begin{pmatrix}0 & -i\\
i & 0
\end{pmatrix}}\label{eq:10}
\end{equation}
Consequently, equation (\ref{eq:9}) transforms into:
\begin{equation}
\left\{ \beta^{0}\xi-\beta^{1}\left(p_{x}-iMB\omega x\right)-\beta^{2}\left(p_{y}-iMB\omega y\right)-M\left(\gamma^{0}\otimes\gamma^{0}\right)\right\} \psi_{k}=0\label{eq:11}
\end{equation}
Utilizing the following relationships:
\begin{equation}
\beta^{0}=\gamma^{0}\otimes I+I\otimes\gamma^{0}\label{eq:12}
\end{equation}
\begin{equation}
\beta^{1}=\gamma^{1}\otimes\hat{I}+\hat{I}\otimes\gamma^{1}=\hat{I}\otimes\gamma^{0}\sigma_{x}+\gamma^{0}\sigma_{x}\otimes\hat{I}\label{eq:13}
\end{equation}
\begin{equation}
\beta^{2}=\hat{I}\otimes\gamma^{0}\sigma_{y}+\gamma^{0}\sigma_{y}\otimes\hat{I}\label{eq:14}
\end{equation}
and multiplying equation (\ref{eq:11}) on the left by $\left(\gamma^{0}\otimes\gamma^{0}\right)$,
we derive:
\begin{equation}
\left\{ \left(\gamma^{0}\otimes I+I\otimes\gamma^{0}\right)\xi-\sqcup-\sqcap-M\right\} \psi_{k}=0\label{eq:15}
\end{equation}
where 
\begin{equation}
\sqcup=\left(\gamma^{0}\otimes\sigma_{x}+\sigma_{x}\otimes\gamma^{0}\right)\left(p_{x}-iMB\omega x\right)\label{eq:16}
\end{equation}
 and 
\begin{equation}
\sqcap=\left(\gamma^{0}\otimes\sigma_{x}+\sigma_{x}\otimes\gamma^{0}\right)\left(p_{y}-iMB\omega y\right)\label{eq:17}
\end{equation}
Here, for simplicity, we have set $\hbar=c=1$.

Injecting equation (\ref{eq:8}) into equation (\ref{eq:11}) yields:
\begin{equation}
\begin{array}{c}
\left(2E-M\right)\psi_{1}-\left\{ \left(p_{x}+iM\omega x\right)-i\left(p_{y}+iM\omega y\right)\right\} \psi_{2}-\left\{ \left(p_{x}+iM\omega x\right)-i\left(p_{y}-iM\omega y\right)\right\} \psi_{3}=0\end{array}\label{eq:18}
\end{equation}
\begin{equation}
\begin{array}{c}
M\psi_{2}-\left\{ \left(p_{x}-iM\omega x\right)+i\left(p_{y}-iM\omega y\right)\right\} \psi_{1}+\left\{ \left(p_{x}-iM\omega x\right)-i\left(p_{y}-iM\omega y\right)\right\} \psi_{4}=0\end{array}\label{eq:19}
\end{equation}
\begin{equation}
\begin{array}{c}
M\psi_{3}-\left\{ \left(p_{x}-iM\omega x\right)+i\left(p_{y}-iM\omega y\right)\right\} \psi_{1}+\left\{ \left(p_{x}-iM\omega x\right)-i\left(p_{y}-iM\omega y\right)\right\} \psi_{4}=0\end{array}\label{eq:20}
\end{equation}
\begin{equation}
\begin{array}{c}
-\left(2E+M\right)\psi_{4}+\left\{ \left(p_{x}+iM\omega x\right)+i\left(p_{y}+iM\omega y\right)\right\} \psi_{2}+\left\{ \left(p_{x}+iM\omega x\right)+i\left(p_{y}+iM\omega y\right)\right\} \psi_{3}=0\end{array}\label{eq:21}
\end{equation}
Now, let's transform these equations into the non-commutative (NC)
space.

\subsection{The solutions in NC space}

To initiate, we recognize that the non-commutative phase space is
defined by the relationship between its coordinate operators as detailed
in references {[}15, 16, 17, 18, 19, 20{]}:
\begin{equation}
\left[x_{\nu}^{\left(NC\right)},x_{\mu}^{\left(NC\right)}\right]=i\Theta_{\mu\nu},\quad\left[p_{\mu}^{\left(NC\right)},p_{\nu}^{\left(NC\right)}\right]=i\bar{\Theta}_{\mu\nu},\quad\left[x_{\mu}^{\left(NC\right)},p_{\nu}^{\left(NC\right)}\right]=i\delta_{\mu\nu}\label{eq:22}
\end{equation}
Here, $\Theta_{\mu\nu}$ and $\bar{\Theta}_{\mu\nu}$ represent an
antisymmetric tensor of spatial dimension. To preserve unitarity and
causality in our theory, we set $\Theta_{0\nu}=0$, ensuring that
time remains a parameter unaffected by non-commutativity, which solely
impacts physical space. These non-commutative models, described by
Equation (1), can be realized using a \ensuremath{\star}-product,
where the commutative algebra of functions with the usual product
$f(x)g(x)$ is replaced by the Moyal algebra \cite{13,14}:
\begin{equation}
\left(f\star g\right)\left(x\right)=\exp\left[\frac{i}{2}\tilde{\theta}_{ab}\partial_{x_{a}}\partial_{x_{b}}\right]f\left(x\right)g\left(y\right)\bigg|_{x=y}\label{eq:23}
\end{equation}
Given our system's two-dimensional nature, we confine our examination
to the $xy$ plane, where the non-commutative algebra takes the form:
\begin{equation}
\left[x_{i}^{\left(NC\right)},x_{j}^{\left(NC\right)}\right]=i\Theta\epsilon_{ij},\quad\left[p_{i}^{\left(NC\right)},p_{j}^{\left(NC\right)}\right]=i\bar{\Theta}\epsilon_{ij},\label{eq:24}
\end{equation}
\begin{equation}
\left[x_{i}^{\left(NC\right)},p_{j}^{\left(NC\right)}\right]=i\delta_{ij},\quad(i,j=1,2)\label{eq:25}
\end{equation}
where $\epsilon_{ij}$ denotes the two-dimensional Levi-Civita tensor.
Rather than resolving the non-commutative Kemmer equation via the
star product method, we opt for Bopp's shift approach. This involves
replacing the star product with the ordinary product using a Bopp's
shift:
\begin{equation}
x_{i}^{\left(NC\right)}=x_{i}-\frac{1}{2}\Theta\epsilon_{ij}p_{j},\quad p_{i}^{\left(NC\right)}=p_{i}+\frac{1}{2}\bar{\Theta}\epsilon_{ij}x_{j}\label{eq:26}
\end{equation}
Consequently, in the two-dimensional non-commutative phase space,
Equation (\ref{eq:26}) transforms into:
\begin{equation}
\hat{x}=x-\frac{1}{2}\Theta p_{y},\quad\hat{y}=y+\frac{1}{2}\Theta p_{x},\label{eq:27}
\end{equation}
\begin{equation}
\hat{p}_{x}=p_{x}+\frac{1}{2}\bar{\Theta}y,\quad\hat{p}_{y}=p_{y}-\frac{1}{2}\bar{\Theta}x\label{eq:28}
\end{equation}
In the literature (see Ref. \cite{20} and associated references),
an upper bound on the value of the coordinate commutator is provided
as $\Theta\leq4\times10^{-40}\,\text{m}^{2}$ and $\bar{\Theta}=2.32\times10^{-61}\,\text{kg}^{2}\,\text{m}^{2}\,\text{s}^{-2}$.
As an approximation, all terms involving the square of $\Theta$ and
$\bar{\Theta}$ have been neglected. It's noteworthy that these values
are theoretically estimated. The final expression necessitates numerical
computation for accessing specific values, and the commutative limit
is effectively covered by setting $\Theta=\bar{\Theta}=0$.

By substituting (\ref{eq:26}) and (\ref{eq:28}) into (\ref{eq:18})
to (\ref{eq:21}), we obtain:
\begin{equation}
\begin{array}{c}
\left(2\xi-M\right)\psi_{1}-\left\{ \left(\hat{p}_{x}+iM\omega\hat{x}\right)-i\left(\hat{p}_{y}+iM\omega\hat{y}\right)\right\} \psi_{2}-\left\{ \left(\hat{p}_{x}+iM\omega\hat{x}\right)-i\left(\hat{p}_{y}+iM\omega\hat{y}\right)\right\} \psi_{3}=0\end{array}\label{eq:29}
\end{equation}
\begin{equation}
\begin{array}{c}
M\psi_{2}-\left\{ \left(\hat{p}_{x}-iM\omega\hat{x}\right)+i\left(\hat{p}_{y}-iM\omega\hat{y}\right)\right\} \psi_{1}+\left\{ \left(\hat{p}_{x}-iM\omega\hat{x}\right)-i\left(\hat{p}_{y}-iM\omega\hat{y}\right)\right\} \psi_{4}=0\end{array}\label{eq:30}
\end{equation}
\begin{equation}
\begin{array}{c}
M\psi_{3}-\left\{ \left(\hat{p}_{x}-iM\omega\hat{x}\right)+i\left(\hat{p}_{y}-iM\omega\hat{y}\right)\right\} \psi_{1}+\left\{ \left(\hat{p}_{x}-iM\omega\hat{x}\right)-i\left(\hat{p}_{y}-iM\omega\hat{y}\right)\right\} \psi_{4}=0\end{array}\label{eq:31}
\end{equation}
\begin{equation}
\begin{array}{c}
-\left(2\xi+M\right)\psi_{4}+\left\{ \left(\hat{p}_{x}+iM\omega\hat{x}\right)+i\left(\hat{p}_{y}+iM\omega\hat{y}\right)\right\} \psi_{2}+\left\{ \left(\hat{p}_{x}+iM\omega\hat{x}\right)+i\left(\hat{p}_{y}+iM\omega\hat{y}\right)\right\} \psi_{3}=0\end{array}\label{eq:32}
\end{equation}
Now, utilizing the following Bopp shift relations:
\begin{equation}
\hat{x}=x-\frac{1}{2}\Theta p_{y},\quad\hat{y}=y+\frac{1}{2}\Theta p_{x},\label{eq:33}
\end{equation}
\begin{equation}
\hat{p}_{x}=p_{x}+\frac{1}{2}\bar{\Theta}y,\quad\hat{p}_{y}=p_{y}-\frac{1}{2}\bar{\Theta}x\label{eq:34}
\end{equation}
Eqs. (\ref{eq:29}) to (\ref{eq:32}) become:
\begin{equation}
\begin{array}{c}
\left(2E-M\right)\psi_{1}-\left\{ \Omega\left(p_{x}-ip_{y}\right)+iM\omega\Gamma\left(x-iy\right)\right\} \psi_{2}-\left\{ \Omega\left(p_{x}-ip_{y}\right)+iM\omega\Gamma\left(x-iy\right)\right\} \psi_{3}=0\end{array}\label{eq:35}
\end{equation}
\begin{equation}
\begin{array}{c}
M\psi_{2}-\left\{ \Omega\left(p_{x}+ip_{y}\right)-iM\omega\Gamma\left(x+iy\right)\right\} \psi_{1}+\left\{ \bar{\Omega}\left(p_{x}-ip_{y}\right)-iM\omega\bar{\Gamma}\left(x-iy\right)\right\} \psi_{4}=0\end{array}\label{eq:36}
\end{equation}
\begin{equation}
\begin{array}{c}
M\psi_{3}-\left\{ \Omega\left(p_{x}+ip_{y}\right)-iM\omega\Gamma\left(x+iy\right)\right\} \psi_{1}+\left\{ \bar{\Omega}\left(p_{x}-ip_{y}\right)-iM\omega\bar{\Gamma}\left(x-iy\right)\right\} \psi_{4}=0\end{array}\label{eq:37}
\end{equation}
\begin{equation}
\begin{array}{c}
-\left(2\xi+M\right)\psi_{4}+\left\{ \bar{\Omega}\left(p_{x}+ip_{y}\right)+iM\omega\bar{\Gamma}\left(x+iy\right)\right\} \psi_{2}+\left\{ \bar{\Omega}\left(p_{x}+ip_{y}\right)+iM\omega\bar{\Gamma}\left(x+iy\right)\right\} \psi_{3}=0\end{array}\label{eq:38}
\end{equation}
with
\begin{equation}
\Omega=1+\frac{M\omega\Theta}{2},\quad\bar{\Omega}=1-\frac{M\omega\Theta}{2},\label{eq:39}
\end{equation}
\begin{equation}
\Gamma=1+\frac{\bar{\Theta}}{2M\omega},\quad\bar{\Gamma}=1-\frac{\bar{\Theta}}{2M\omega}\label{eq:40}
\end{equation}
Based on these equations, we will have:
\begin{equation}
\begin{array}{c}
\psi_{1}=\frac{2\cup}{2\xi-M}\psi_{2},\end{array}\label{eq:41}
\end{equation}
\begin{equation}
\psi_{2}=\psi_{3}\label{eq:42}
\end{equation}
\begin{equation}
\psi_{4}=\frac{2\cap}{2\xi+M}\psi_{2}\label{eq:43}
\end{equation}
with 
\begin{equation}
\cup=\Omega\left(p_{x}-ip_{y}\right)+iM\omega\Gamma\left(x-iy\right),\quad\cup'=\Omega\left(p_{x}+ip_{y}\right)-iM\omega\Gamma\left(x+iy\right)\label{eq:44}
\end{equation}
\begin{equation}
\cap=\bar{\Omega}\left(p_{x}+ip_{y}\right)+iM\omega\bar{\Gamma}\left(x+iy\right),\quad\cap'=\bar{\Omega}\left(p_{x}-ip_{y}\right)-iM\omega\bar{\Gamma}\left(x-iy\right)\label{eq:45}
\end{equation}
Substituting these equations into (\ref{eq:37}) yields:
\begin{equation}
\left(M-\frac{2\cup'\cup}{2\xi-M}+\frac{2\cap'\cap}{2\xi+M}\right)\psi_{2}\left(x,y\right)=0\label{eq:46}
\end{equation}
where
\begin{equation}
\cup'\cup=\Omega^{2}\left(p_{x}^{2}+p_{y}^{2}\right)+M^{2}\omega^{2}\Gamma^{2}\left(x^{2}+y^{2}\right)+2M\omega\Gamma\Omega-2M\omega\Gamma\Omega L_{z},\label{eq:47}
\end{equation}
\begin{equation}
\cap'\cap=\bar{\Omega}^{2}\left(p_{x}^{2}+p_{y}^{2}\right)+M^{2}\omega^{2}\bar{\Gamma}^{2}\left(x^{2}+y^{2}\right)+2M\omega\bar{\Gamma}\bar{\Omega}+2M\omega\bar{\Gamma}\bar{\Omega}L_{z}.\label{eq:48}
\end{equation}
Upon substituting these expressions into equation (\ref{eq:46}),
we obtain:
\begin{equation}
\begin{aligned} & \left\{ \delta\left(p_{x}^{2}+p_{y}^{2}\right)+\delta'M^{2}\omega^{2}\left(x^{2}+y^{2}\right)+2\beta''M\omega+2\gamma''M\omega L_{z}-\frac{M}{2}\right\} \psi_{2}=0.\end{aligned}
\label{eq:49}
\end{equation}
Here,
\begin{equation}
\delta=\frac{\left(2\xi+M\right)\Omega^{2}-\left(2\xi-M\right)\bar{\Omega}^{2}}{\left(2\xi\right)^{2}-M^{2}},\label{eq:50}
\end{equation}
\begin{equation}
\delta'=\frac{\left(2\xi+M\right)\Gamma^{2}-\left(2\xi-M\right)\bar{\Gamma}^{2}}{\left(2\xi\right)^{2}-M^{2}},\label{eq:51}
\end{equation}
\begin{equation}
\beta''=\frac{\left(2\xi+M\right)\Gamma\Omega-\left(2\xi-M\right)\bar{\Gamma}\bar{\Omega}}{\left(2\xi\right)^{2}-M^{2}},\label{eq:52}
\end{equation}
\begin{equation}
\gamma''=\frac{\left(2\xi+M\right)\Gamma\Omega+\left(2\xi-M\right)\bar{\Gamma}\bar{\Omega}}{\left(2\xi\right)^{2}-M^{2}}.\label{eq:53}
\end{equation}
Equation (\ref{eq:49}) can be reformulated in another manner as:
\begin{equation}
\left\{ \frac{\left(p_{x}^{2}+p_{y}^{2}\right)}{2M}+\frac{1}{2}M\left(\frac{\delta'}{\delta}\omega^{2}\right)\left(x^{2}+y^{2}\right)+\omega\frac{\beta''}{\delta}-\omega L_{z}\frac{\gamma''}{\delta}-\frac{1}{4\delta}\right\} \psi_{2}=0,\label{eq:54}
\end{equation}
with
\begin{equation}
\frac{\delta'}{\delta}=\frac{\left(2\xi+M\right)\Gamma^{2}-\left(2\xi-M\right)\bar{\Gamma}^{2}}{\left(2\xi+M\right)\Omega^{2}-\left(2\xi-M\right)\bar{\Omega}^{2}}=\frac{M+\frac{2\xi\bar{\Theta}}{M\omega}}{M+\xi M\omega\Theta},\label{eq:55}
\end{equation}
\begin{equation}
\frac{\beta''}{\delta}=\frac{\left(2\xi+M\right)\Gamma\Omega-\left(2\xi-M\right)\bar{\Gamma}\bar{\Omega}}{\left(2\xi+M\right)\Omega^{2}-\left(2\xi-M\right)\bar{\Omega}^{2}}=\frac{2M+M^{2}\omega\Theta+\frac{\bar{\Theta}}{\omega}}{2M+2\xi M\omega\Theta},\label{eq:56}
\end{equation}
\begin{equation}
\frac{\gamma''}{\delta}=\frac{\left(2\xi+M\right)\Gamma\Omega+\left(2\xi-M\right)\bar{\Gamma}\bar{\Omega}}{\left(2\xi+M\right)\Omega^{2}-\left(2\xi-M\right)\bar{\Omega}^{2}}=\frac{4\xi+M^{2}\omega\Theta+\frac{\bar{\Theta}}{\omega}}{2M+2\xi M\omega\Theta},\label{eq:57}
\end{equation}
\begin{equation}
\frac{1}{\delta}=\frac{\left(2\xi\right)^{2}-M^{2}}{\left(2\xi+M\right)\Omega^{2}-\left(2\xi-M\right)\bar{\Omega}^{2}}=\frac{\left(2\xi\right)^{2}-M^{2}}{2M+2\xi M\omega\Theta},\label{eq:58}
\end{equation}
where
\begin{equation}
\omega'=\sqrt{\frac{\delta'}{\delta}}\omega=\sqrt{\frac{M^{2}\omega+2\bar{\Theta}\xi}{M+2M\omega\Theta\xi}}\omega.\label{eq:59}
\end{equation}
Concerning the eigensolutions of our given problem, we encounter the
following differential equation:
\begin{equation}
\left\{ \frac{\left(p_{x}^{2}+p_{y}^{2}\right)}{2M}+\frac{1}{2}M\omega'{}^{2}\left(x^{2}+y^{2}\right)\right\} \psi_{2}=\tilde{\xi}\psi_{2}.\label{eq:60}
\end{equation}
with
\begin{equation}
\tilde{\xi}=-\omega L_{z}\frac{\gamma''}{\delta}-\omega\frac{\beta''}{\delta}+\frac{1}{4\delta}.\label{eq:61}
\end{equation}
Equation (\ref{eq:60}) represents the differential equation governing
the component $\psi_{2}$, enabling us to derive the energy spectrum
of the system under consideration.

\subsection{The solutions in NC space in the presence of a minimal length }

In the minimal length formalism, the Heisenberg algebra is given by
\cite{23,24,25,26}:
\begin{equation}
\left[\hat{x}_{i},\hat{p}_{i}\right]=i\delta_{ij}\left(1+\beta p^{2}\right)\label{eq:78}
\end{equation}
where $0<\beta\leq1$ is the minimal length parameter and $p$ is
the magnitude of the momentum. When the energy is much smaller than
the Planck mass, $\beta$ goes to zero and we recover the Heisenberg
uncertainty principle. A representation of $\hat{x}_{i}$ and $\hat{p}_{i}$
satisfying Eq. (\ref{eq:78}) may be chosen as
\begin{equation}
\hat{x}=i\left(1+\beta p^{2}\right)\frac{d}{dp_{x}}=i\left(1+\beta p^{2}\right)\left(\cos\theta\frac{d}{dp}-\frac{\sin\theta}{p}\frac{d}{d\theta}\right),\,\hat{p}_{x}=p_{x},\label{eq:79}
\end{equation}
\begin{equation}
\hat{y}=i\left(1+\beta p^{2}\right)\frac{d}{dp_{y}}=i\left(1+\beta p^{2}\right)\left(\sin\theta\frac{d}{dp}+\frac{\cos\theta}{p}\frac{d}{d\theta}\right),\,\hat{p}_{y}=p_{y},\label{eq:80}
\end{equation}
with $p^{2}=p_{x}^{2}+p_{y}^{2}$. By using the equations (\ref{eq:79})
and (\ref{eq:80}), Eq. (\ref{eq:60}) becomes 
\begin{equation}
\left[p^{2}-M^{2}\omega'{}^{2}\left(1+\beta p^{2}\right)^{2}\left(\frac{\partial^{2}}{\partial p^{2}}+\frac{1}{p}\frac{\partial}{\partial p}+\frac{1}{p^{2}}\frac{\partial^{2}}{\partial\theta^{2}}\right)+2M\omega L_{z}\frac{\gamma^{''}}{\delta}+2M\omega\frac{\beta^{''}}{\delta}-\frac{M}{2\delta}\right]\varPsi_{2}=0.\label{eq:81}
\end{equation}
\foreignlanguage{ngerman}{Now, when we put that 
\begin{equation}
\psi_{2}=h(p)e^{i\left|j\right|\theta},\label{eq:82}
\end{equation}
with $j=0,\pm1,\pm2,\ldots$, the Eq. (\ref{eq:81}) is transformed
into 
\begin{equation}
\left\{ -a(p)\frac{\partial^{2}}{\partial p^{2}}+b(p)\frac{\partial}{\partial p}+c(p)-\varsigma^{2}\right\} h(p)=0,\label{eq:83}
\end{equation}
with 
\begin{equation}
\varsigma^{2}=-2M\omega\left|j\right|\frac{\gamma^{''}}{\delta}-2M\omega\frac{\beta^{''}}{\delta}+\frac{M}{2\delta}\label{eq:84}
\end{equation}
\begin{equation}
\begin{array}{c}
a\left(p\right)=M^{2}\omega'{}^{2}\left(1+\beta p^{2}\right)^{2},\,b\left(p\right)=-\frac{M^{2}\omega'{}^{2}\left(1+\beta p^{2}\right)^{2}}{p}\\
c(p)=p^{2}+\frac{M^{2}\omega'{}^{2}\left(1+\beta p^{2}\right)^{2}j^{2}}{p^{2}}
\end{array},\label{eq:86}
\end{equation}
}To address Eq. (\ref{eq:83}), we employ a substitution pattern utilized
by Jana et al. \cite{45}, where
\begin{equation}
h(p)=\rho(p)\varphi(p),\,q=\int\frac{1}{\sqrt{a(p)}}dp\label{eq:87}
\end{equation}
 Here, 
\begin{equation}
\rho(p)=\exp\left(\int\chi(p)dp\right)\label{eq:88}
\end{equation}
with 
\begin{equation}
\chi(p)=\frac{2b+a'}{4a}=-\frac{1}{2p}+\frac{\beta p}{1+\beta p^{2}}\label{eq:89}
\end{equation}
This substitution yields 
\begin{equation}
\rho(p)=\sqrt{\frac{1+\beta p^{2}}{p}}\label{eq:90}
\end{equation}
After these substitutions, Eq. (\ref{eq:83}) transforms to 
\begin{equation}
\left[-\frac{d^{2}\varphi(p)}{dq^{2}}+V(p)\right]\varphi(p)=\varsigma\varphi(p)\label{eq:91}
\end{equation}
where 
\begin{equation}
V(p)=p^{2}+\frac{M^{2}\omega'{}^{2}\left(1+\beta p^{2}\right)^{2}j^{2}}{p^{2}}-\frac{M^{2}\omega'{}^{2}}{4p^{2}}-\frac{M^{2}\omega'{}^{2}\beta\left(1+\beta p^{2}\right)}{4}\label{eq:92}
\end{equation}
To simplify $V(p)$, we introduce a change of variable $p=\frac{1}{\sqrt{\beta}}\tan\left(q\lambda\sqrt{\beta}\right)$,
where $\lambda=M^{2}\omega'{}^{2}$. With this transformation, $V(p)$
becomes 
\begin{equation}
\ensuremath{V(q)=-\frac{1}{\beta}+\frac{\lambda^{2}\beta}{4}+\underbrace{\beta\lambda^{2}}_{U_{0}}\left(\frac{\frac{1}{\beta^{2}\lambda^{2}}+j^{2}-\frac{1}{4}}{\cos^{2}\alpha q}+\frac{j^{2}-\frac{1}{4}}{\sin^{2}\alpha q}\right)}.\label{eq:93}
\end{equation}
Consequently, the final form of our differential equation becomes
\begin{equation}
\ensuremath{\left\{ -\frac{d^{2}\varphi(p)}{dq^{2}}+U_{0}\left(\frac{\frac{1}{\beta^{2}\lambda^{2}}+j^{2}-\frac{1}{4}}{\cos^{2}\alpha q}+\frac{j^{2}-\frac{1}{4}}{\sin^{2}\alpha q}\right)\right\} \varphi(p)=\bar{\varsigma}^{2}\varphi(p)}\label{eq:94}
\end{equation}
where
\begin{equation}
\ensuremath{\bar{\varsigma}^{2}=\frac{1}{\beta}-\frac{\lambda^{2}\beta}{4}-2M\omega\left|j\right|\frac{\gamma^{''}}{\delta}-2M\omega\frac{\beta^{''}}{\delta}+\frac{M}{2\delta}}\label{eq:95}
\end{equation}
Thus, Eq. (\ref{eq:94}) can be represented as 
\begin{equation}
\ensuremath{\left\{ -\frac{d^{2}\varphi(p)}{dq^{2}}+U_{0}\left(\frac{\zeta_{1}\left(\zeta_{1}-1\right)}{\cos^{2}\alpha q}+\frac{\zeta_{2}\left(\zeta_{2}-1\right)}{\sin^{2}\alpha q}\right)\right\} \varphi(q)=\bar{\varsigma}^{2}\varphi(q)},\label{eq:96}
\end{equation}
Here
\begin{equation}
\zeta_{1}\left(\zeta_{1}-1\right)=\frac{1}{\beta^{2}\lambda^{2}}+j^{2}-\frac{1}{4},\label{eq:97}
\end{equation}
\begin{equation}
\zeta_{2}\left(\zeta_{2}-1\right)=j^{2}-\frac{1}{4}\label{eq:98}
\end{equation}
Finally, we arrive at 
\begin{equation}
\left(-\frac{d^{2}}{dq^{2}}+U_{0}\left\{ \frac{\zeta_{1}\left(\zeta_{1}-1\right)}{\sin^{2}\left(\alpha q\right)}+\frac{\zeta_{2}\left(\zeta_{2}-1\right)}{\cos^{2}\left(\alpha q\right)}\right\} \right)\varphi\left(q\right)=\bar{\varsigma}^{2}\varphi\left(q\right)\label{eq:99}
\end{equation}
where $U_{0}=\alpha^{2}$ and $\alpha=\lambda\sqrt{\beta}$. At this
stage, Eq. (\ref{eq:99}) manifests as the well-known Schrödinger
equation within a Pöschl-Teller potential, characterized by the potential
{[}39{]}:
\begin{equation}
U=U_{0}\left\{ \frac{\zeta_{1}\left(\zeta_{1}-1\right)}{\sin^{2}\left(\alpha q\right)}+\frac{\zeta_{2}\left(\zeta_{2}-1\right)}{\cos^{2}\left(\alpha q\right)}\right\} \label{eq:100}
\end{equation}
Here, we stipulate the conditions $(\zeta_{1},\zeta_{2})>1$. From
Eqs. (\ref{eq:97}) and (\ref{eq:98}), we deduce:
\begin{equation}
\zeta_{1}=\frac{1}{2}\pm\sqrt{j^{2}+\frac{1}{\beta^{2}\lambda^{2}}}\label{eq:101}
\end{equation}
\begin{equation}
\zeta_{2}=\left|j\right|\pm\frac{1}{2}\label{eq:102}
\end{equation}
To solve Eq. (\ref{eq:99}), we introduce the novel variable $z=\sin^{2}\left(\alpha q\right)$.
Consequently, Eq. (\ref{eq:99}) can be reformulated as:
\begin{equation}
z\left(1-z\right)\varphi^{''}+\left(\frac{1}{2}-z\right)\varphi^{'}+\frac{1}{4}\left\{ \frac{\bar{\varsigma}^{2}}{\alpha^{2}}-\frac{\zeta_{1}\left(\zeta_{1}-1\right)}{z}-\frac{\zeta_{2}\left(\zeta_{2}-1\right)}{1-z}\right\} \varphi=0\label{eq:103}
\end{equation}
With the following Ansatz on the wave function $\varphi$, defined
as \cite{46}
\begin{equation}
\varphi=z^{\frac{\zeta_{1}}{2}}\left(1-z\right)^{\frac{\zeta_{2}}{2}}\Psi\left(z\right),\label{eq:104}
\end{equation}
we obtain:
\begin{equation}
z\left(1-z\right)\Psi^{''}+\left[\left(\zeta_{1}+\frac{1}{2}\right)-z\left(\zeta_{1}+\zeta_{2}+1\right)\right]\Psi^{'}+\frac{1}{4}\left\{ \frac{\bar{\xi}}{\alpha}-\left(\zeta_{1}+\zeta_{2}\right)^{2}\right\} \Psi=0\label{eq:105}
\end{equation}
The general solution to this equation is given by:
\begin{equation}
\Psi=C_{1}\,_{2}F_{1}\left(a';b';c';z\right)+C_{2}\,z^{1-c}\,_{2}F_{1}\left(a'+1-c';b'+1-c';2-c;z\right)\label{eq:106}
\end{equation}
Where 
\begin{equation}
a'=\frac{1}{2}\left(\zeta_{1}+\zeta_{2}+\frac{\bar{\varsigma}}{\alpha}\right),\,b'=\frac{1}{2}\left(\zeta_{1}+\zeta_{2}-\frac{\bar{\varsigma}}{\alpha^{2}}\right),\,c'=\zeta_{1}+\frac{1}{2}.\label{eq:107}
\end{equation}
The second solution displays singularities at $z=0$ and $z=1$. Consequently,
we exclude it from consideration, focusing solely on the first term
in the solutions (refer to pages 90-92 in Ref. \cite{46}). The confluent
series transforms into a polynomial if and only if $a=-n$, where
$n=0,1,2,\ldots$. With this condition, we proceed to derive...
\begin{equation}
\bar{\varsigma}^{2}=\alpha^{2}\left(\zeta_{1}+\zeta_{2}+2n\right)^{2}\label{eq:108}
\end{equation}
where, the precise expressions for $\zeta_{1}$ and $\zeta_{2}$,
satisfying the condition $(\zeta_{1},\zeta_{2})>1$, are
\begin{equation}
\zeta_{1}=\frac{1}{2}+\sqrt{j^{2}+\frac{1}{\beta^{2}\lambda^{2}}},\label{eq:109}
\end{equation}
\begin{equation}
\zeta_{2}=\left|j\right|+\frac{1}{2}.\label{eq:110}
\end{equation}
with $j\neq0$. Utilizing Eqs. (\ref{eq:56}), (\ref{eq:57}), and
(\ref{eq:58}) with (\ref{eq:108}), (\ref{eq:109}) and (\ref{eq:110}),
we derive the final expression for the energy spectrum:
\begin{equation}
\begin{aligned}
& \frac{\left(2\xi\right)^{2}-M^{2}}{4+4\xi M\omega\Theta}-2M\omega\left|j\right|\left(\frac{4\xi+M^{2}\omega\Theta+\frac{\bar{\Theta}}{\omega}}{2M+2\xi M\omega\Theta}\right)\\
& -2M\omega\left(\frac{2M+M^{2}\omega\Theta+\frac{\bar{\Theta}}{\omega}}{2M+2\xi M\omega\Theta}\right)\\
& =-\frac{1}{\beta}+\beta\left(\frac{M^{3}\omega^{3}+2M\omega^{2}\bar{\Theta}\xi}{1+2\omega\Theta\xi}\right)\left(\frac{1}{2}+\left|j\right|+\sqrt{j^{2}+\frac{1}{\beta^{2}\lambda^{2}}}+2n\right)^{2}
\end{aligned}
\label{eq:111}
\end{equation}
Equation (\ref{eq:111}) presents a numerical expression for the energy
spectrum. These energies can be computed efficiently using the current
numerical techniques. When $\bar{\Theta}=\Theta=0$, the solutions
revert to those in the commutative space. Notably, parameters such
as $\beta$, $\Theta$, and $\bar{\Theta}$ introduce variations in
the energy spectrum, breaking its degeneracy. Furthermore, the presence
of $\beta$ in the energy spectrum, unlike $\Theta$ and $\bar{\Theta}$,
causes the energy levels to depend on $n^{2}$, a characteristic indicative
of hard confinement, as indicated by Nouicer et al. \cite{39,40,41}.

The final wave function is then
\begin{equation}
\hat{\psi}_{K}\left(x,y\right)=C_{1}\left(\begin{array}{c}
\frac{\Omega\left(p_{x}-ip_{y}\right)+iM\omega\Gamma\left(x-iy\right)}{2\xi-M}\\
1\\
1\\
\frac{\bar{\Omega}\left(p_{x}+ip_{y}\right)+iM\omega\bar{\Gamma}\left(x+iy\right)}{2\xi+M}
\end{array}\right)e^{i\left|j\right|\theta}\sqrt{\frac{1+\beta p^{2}}{p}}z^{\frac{\zeta_{1}}{2}}\left(1-z\right)^{\frac{\zeta_{2}}{2}}\,_{2}F_{1}\left(a';b';c';z\right).\label{eq:112}
\end{equation}

\section{Conclusion}

In this paper, we present an exact solution to the Kemmer oscillator
in two dimensions within the framework of relativistic quantum mechanics,
considering both minimal length and non-commutative (NC) space effects.
Subsequently, we introduce the concept of minimal length into our
analysis. This is accomplished by: (i) incorporating the coordinates
of non-commutative space with those of commutative space using the
Bopp shift approximation, and (ii) subsequently integrating the minimal
length into our equations. This augmentation transforms the problem
into one featuring a Pöschl-Teller potential.

The pronounced dependence of these solutions on both the minimal length
and non-commutative parameters is evident. Furthermore, a comparative
analysis of the energy spectrum derived in our study with that of
the same problem in flat space-time reveals the disruptive influence
of both parameters on the energy spectrum's degeneracy. Notably, the
emergence of a term proportional to $n^{2}$ in the energy spectrum
indicates the presence of hard confinement, a feature distinctly manifest
upon the introduction of minimal length into our analysis.

\end{document}